\documentclass[12pt]{iopart}
\usepackage{graphicx}
\begin{document}

\title{Direct and inverse solver for the 3D optoacoustic Volterra equation}

\author{J Stritzel, O Melchert, M Wollweber and B Roth}

\address{Hannover Centre for Optical Technologies (HOT), Leibniz Universit\"at Hannover, Nienburger Str.\,17, D-30167 Hannover, Germany}
\ead{jenny.stritzel@hot.uni-hannover.de}

\begin{abstract}
The {\emph{direct problem}} of optoacoustic signal generation in biological
media consists of solving the inhomogeneous optoacoustic wave equation for an
initial acoustic stress profile. In contrast, the mathematically challenging
{\emph{inverse problem}} requires the reconstruction of the initial stress 
profile from a proper set of observed signals.

In this article, we consider the particular case of a Gaussian transverse
irradiation source profile in the paraxial approximation of the wave equation,
for which the direct problem along the beam axis can be cast into a linear
Volterra integral equation of the second kind.  This integral equation can be
used in two ways: as a {\emph{forward solver}} to predict optoacoustic signals
in terms of the direct problem, and as an {\emph{inverse solver}} for which we
here devise highly efficient numerical schemes used for the reconstruction of
initial pressure profiles from observed signals, constituting a methodical
progress of computational aspects of optoacoustics.

In this regard, we explore the validity as well as the limits of the inversion
scheme via numerical experiments, with parameters geared towards actual
optoacoustic problem instances. The considered inversion input consists of
{\emph{synthetic}} data, obtained by means of forward solvers based on the
Volterra integral, and, more generally, the optoacoustic Poisson integral.
Regarding the latter, we numerically invert signals that correspond to
different detector-to-sample distances and assess the convergence to the true
initial stress profiles upon approaching the far-field.  Finally, we also
address the effect of noise on the quality of the reconstructed pressure
profiles. 
\end{abstract}

%
\vspace{2pc}

%
%
%

\section{Introduction}

The {\emph{inverse}} optoacoustic (OA) source reconstruction problem is
concerned with the recovery of initial acoustic stress profiles from measured
OA signals upon knowledge of the mathematical model that mediates the
underlying diffraction
transformation~\cite{Kuchment:2008,Wang:2009,Colton:2013}.  It is the
conceptual analogue of the {\emph{direct}} OA problem (known as the OA
{\emph{forward}} problem), representing the calculation of a
diffraction-transformed acoustic pressure signal at a given field point for a
given initial acoustic stress profile on the basis of the OA wave equation
\cite{Wang:2009,Colton:2013,Gusev:1993,Landau:1987}.

During the last decades, the former problem has received much attention within
the field of optoacoustics, owing to its immediate relevance for 
medical applications, see, e.g., Refs.\,\cite{Stoffels:2015,Stoffels:2015ERR},
that aim at the reconstructions of ``internal'' OA material properties from
``external'' measurements.
In this regard, current progress is
mostly due to photoacoustic tomography (PAT) and imaging applications supporting 
different approaches that might be divided into three groups: (i)
back-projection approaches in time or frequency domain
\cite{Xu:2005,Kostli:2001}, (ii) time-reverse evolution of the linear
(``T-symmetric'') OA wave equation \cite{Xu:2004,Burgholzer:2007}, (iii)
model-based least-squares schemes \cite{Paltauf:2002b,DeanBen:2012}, also
involving more hybrid approaches, based, e.g., on the ideas of image
reconstruction via compressed sensing \cite{Provost:2009} or Landweber
iteration schemes \cite{Belchami:2016}. All these approaches are equipped with 
their own benefits and drawbacks.
In particular, note that the inversion input for PAT backpropagation approaches
consits of a multitude of signals recorded on a surface enclosing the OA source
volume.  In contrast, we here introduce an alternate
approach that allows for the numerical reconstruction of initial stress
profiles via inversion of signals from ``single-shot'' measurements.
Therefore we focus on the direct and inverse problem in the paraxial
approximation to the full OA wave equation \cite{Karabutov:1998,Gusev:1993},
where we allude to a numerical treatment of an underlying integral equation,
capturing the diffraction-transformation of signals for an on-axis setting,
leading to highly efficient forward and inverse solvers for the OA problem in
the considered setting. 

After developing and testing the numerical procedure, we assess how well the
particular source reconstruction problem performs beyond the paraxial
approximation by considering: (i)~signals obtained for the full OA
wave-equation in the acoustic far-field, and, (ii)~synthetic signals
exhibiting noise.
To the knowledge of the authors, within the field of optoacoustics no such
numerical procedure has been discussed and put under scrutiny, yet.  Hence, we
present a methodical progress in the field of optoacoustics, appealing from
the point of view of computational theoretical physics.

The remainder of the article is organized as follows.
In section \ref{sec:Theo}, we briefly summarize the theoretical background of 
OA signal generation and calculation.
In section \ref{sect:Numerics} we elaborate on our numerical approaches
that yield forward and inverse solvers for the OA problem in the paraxial 
approximation, followed by a sequence of numerical experiments described in 
section \ref{sect:NumEx}. We summarize and conclude upon our
findings in section \ref{sect:Summary}.


\section{Theoretical aspects of optoacoustic (OA) signal generation}
\label{sec:Theo}

In the subsequent subsections we briefly review the theoretical foundation of
the mechanism of optoacoustic signal generation. In this regard, in subsection\
\ref{subsec:Theo_OAPoissonInt}, we first detail a forward solution of the
general problem based on the \emph{OA Poisson integral}. After this, in
subsection\ \ref{subsec:Theo_OAVolterraInt}, we allude to a particular
``on-axis'' variant of the forward problem, paving the way for a highly
efficient inversion scheme in terms of an \emph{OA Volterra integral}.

\subsection{General optoacoustic signal generation -- The OA Poisson integral}
\label{subsec:Theo_OAPoissonInt}

Albeit there are several microscopic mechanisms that possibly contribute to the
generation of optoacoustic signals \cite{Tam:1986}, we restrict the subsequent
theoretical discussion to the most dominant photothermal heating effect, i.e.\
thermal expansion.  Further, we consider a pulsed optoacoustic working mode
with a pulse duration that is:
(i)~significantly smaller than the thermal relaxation time of the surrounding
material \cite{Kruger:1995,comment:ThermalConfinement}, realizing what is 
referred to as \emph{thermal confinement}, and,
(ii)~short enough to be represented by a delta-function on the scale of typical
acoustic propagation times \cite{comment:StressConfinement}, denoted by
\emph{stress confinement}.

Then, focusing on the acoustic aftermath of the thermoelastic expansion 
mechanism, the scalar excess pressure field $p(\vec{r},t)$ at time $t$ and
field point $\vec{r}$ can be related to an initial stress distribution 
$p_0(\vec{r})$ via the inhomogeneous optoacoustic wave equation 
\cite{Wang:2009,Tam:1986} 
\begin{equation}
\big[ \partial_t^2 - c^{2} \vec{\nabla}^2 \big]~p(\vec{r},t) = \partial_t~p_0(\vec{r})\,\delta(t), \label{eq:OAWaveEq}
\end{equation}
wherein $c$ signifies the speed of sound and where the initial stress field is
related to the volumetric energy density $W(\vec{r})$ \cite{Wang:2009,Paltauf:2000}, deposited in the
irradiated region via absorption and photothermal heating by the short laser pulse, according
to $p_0(\vec{r}) = \Gamma W(\vec{r})$. Therein, $\Gamma$ refers to the
Gr\"uneisen parameter, an effective parameter describing the fraction of
absorbed heat that is actually converted to mechanical stress.  Note that in
Eq.\ (\ref{eq:OAWaveEq}), temporal changes of the local photothermal heat
absorption field $W(\vec{r})\delta(t)$ trigger stress waves that propagate
through the medium and constitute the optoacoustic signal.

An analytic solution that yields the excess pressure $p(\vec{r},t)$ according 
to Eq.\,(\ref{eq:OAWaveEq}) is accessible through the corresponding 
Greens-function in free space, establishing the optoacoustic Poisson integral
\cite{Landau:1987,Wang:2009,Kruger:1995}
\begin{equation}
p(\vec{r},t) = \frac{1}{4\pi c} \partial_t
\int\limits_{V}\!\frac{p_0(\hat{\vec{r}})}{|\vec{r}-\hat{\vec{r}}|}
\delta(|\vec{r}-\hat{\vec{r}}| - ct)\,\mathrm{d}\hat{\vec{r}}
,\label{eq:OAPoissonInt}
\end{equation} 
wherein $V$ represents the ``source volume'' beyond which the initial stress
$p_0(\vec{r})=0$ \cite{comment:PrincipalValue}, and $\delta(\cdot)$
limiting the integration to a time-dependent surface centered at $\vec{r}$ and
radially constraint by $|\vec{r}-\hat{\vec{r}}| = ct$.

%
%
\begin{figure}[th!]
\centerline{\includegraphics[width=0.8\linewidth]{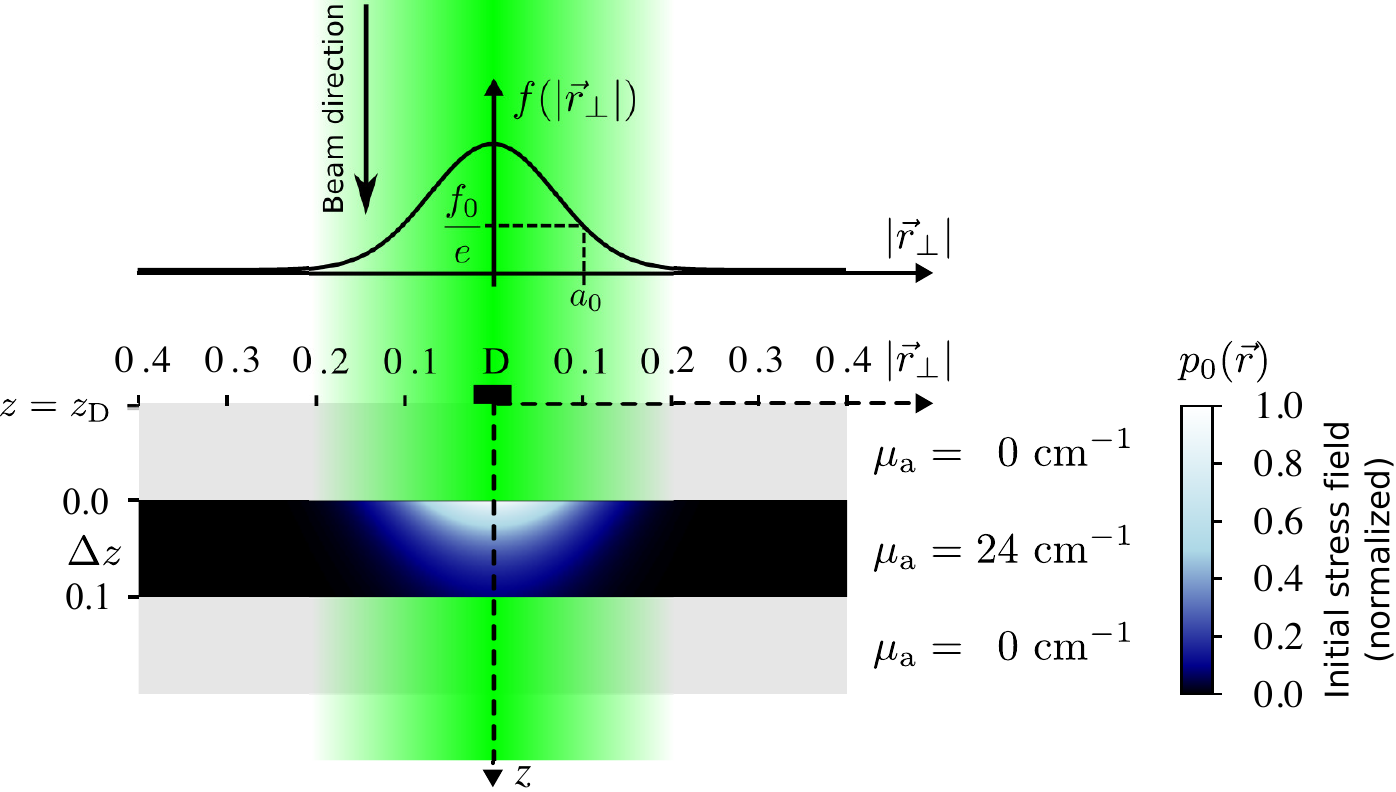} } 
\caption{(Color online) Illustration of an optoacoustic setup with
plane-normal irradiation source profile $f(|\vec{r}_\perp|)$ and a source
volume consisting of possibly multiple stacked absorbing layers. The
irradiation source exhibits a $1/e$--intensity radius of $|\vec{r}_\perp|=a_0$
and the different layers possess absorption coefficients $\mu_a$ as indicated
in the figure.  Here, the figure shows a single absorbing layer of width
$\Delta z = 0.1\,\rm{cm}$.  The initial stress field $\propto p_0(\vec{r})$
causes acoustic pressure waves that can be monitored as an optoacoustic signal
at the detection point ${\rm D}$, here located on the beam axis, i.e.\ at
$|\vec{r}_\perp|=0$, with distance $z_{\rm D}<0$ from the absorbing layer.}
\label{fig:setup}
\end{figure}

\subsection{Diffraction transformation in the paraxial approximation -- The OA Volterra integral equation}
\label{subsec:Theo_OAVolterraInt}

In the remainder of the article we consider non-scattering compounds which
consist of plane-parallel layers, stacked along the $z$-direction of an
associated coordinate system within the positive half space. The acoustic
properties within the bulk are assumed to be constant, whereas the optical
properties are set to be constant within the layers but may differ from layer
to layer, characterized by a depth-dependent absorption coefficient $\mu_{\rm
a}(\vec{r})\equiv\mu_{\rm a}(z)$, see Fig.\ \ref{fig:setup}.  
Then, for an inherently $2{\rm D}$ plane-normal irradiation source profile
$f(\vec{r}_\perp)$, the initial stress field can be factored according to 
\begin{equation}
p_0(\vec{r})~=~\Gamma f(\vec{r}_\perp)\, g(z), \label{eq:iniStress}
\end{equation}
wherein $g(z)$ summarizes the effect of the absorptive properties of the
layered medium in terms of a $1{\rm D}$ axial absorption depth profile.
Bearing in mind that we consider non-scattering media, the latter follows
Beer-Lamberts law, i.e.\
\begin{equation}
g(z) = \mu_{\rm a}(z) \exp\Bigg\{-\int\limits_0^z \mu_{\rm
a}(z^{\prime})~dz^{\prime}\Bigg\}. \label{eq:g}
\end{equation}
Note that such a factorization of $p_0(\vec{r})$ is well justified: there
are numerous studies where experiments and their complementing simulations
are in accord with the above constraints, see, e.g., Refs.\
\cite{Paltauf:1997,Paltauf:1998,Paltauf:2000,Jaeger:2005,Blumenroether:2016}.

Further, let $f(\vec{r}_\perp)$ be an axially symmetric irradiation source
profile with a Gaussian transverse profile, i.e.\
\begin{equation}
f(\vec{r}_\perp) = f_0\, \exp\Big\{-|\vec{r}_\perp|^2/a_0^2\Big\}, \label{eq:isp} 
\end{equation}
wherein $f_0$ signifies the incident radiant exposure on the beam axis
$|\vec{r}_\perp|=0$ and $a_0$ defines the $1/e$-threshold of the beam intensity.
This is a realistic assumption for many excitation sources applied in OA. 

Under the above prerequisites, it can be shown that the diffraction
transformation of laser-excited excess pressure profiles in the paraxial
approximation of Eq.~(\ref{eq:OAWaveEq}) at a detection point $\vec{r}_{\rm
D}$ along the beam axis, i.e.\
\mbox{$p_{\rm D}(\tau)\equiv p(\vec{r}_{\rm D},t)|_{(\vec{r}_\perp=0,
\tau=t+z_{\rm D}/c)}$}, can properly be described in terms of a Volterra
integral equation of the $2$nd kind \cite{Gusev:1993,Karabutov:1998,comment:onAxis}, here referred to as the \emph{optoacoustic
Volterra integral}, reading
\begin{equation}
\label{Volteq}
p_{\rm D}(\tau) = p_0(\tau) - \int\limits_{-\infty}^\tau \!\mathcal{K}(t,\tau)\,p_0(t)\,\mathrm{d}t.\label{eq:OAVolterraInt}
\end{equation}
In the above, the change of argument in the description of the initial stress
has to be understood as $p_0(\tau) =
p_0(\vec{r})|_{(\vec{r}_\perp=0,\tau=t+z_{\rm D}/c)}$.  The Volterra operator,
i.e.\ the second term in the equation above, describes the diffraction 
transformation experienced by the OA signal.
It governs the propagation of acoustic
stress waves in the optoacoustic on-axis setting with Gaussian irradiation source
profile via a convolution type Volterra kernel 
$\mathcal{K}(t,\tau)\,=\,\mathcal{K}(\tau - t)$, 
wherein \cite{Karabutov:1998}
\begin{equation}
\mathcal{K}(\tau - t) =\omega_{\rm D}\, \exp \big\{-\omega_{\rm D}\,(\tau - t)\big\}. \label{eq:OAPropagator}
\end{equation}
Therein $\omega_{\rm D} = 2 c |z_{\rm D}|/ a_0^2$ denotes a characteristic OA
frequency, related to the two ``exterior'' OA lengthscales given by: (i) the
distance $|z_{\rm D}|$ between detection point and absorbing layer, and, (ii)
the transversal characteristic lengthscale $a_0$ of the irradiation source
profile. 
The dependence of the diffraction transformation on the 
frequencies $\omega_{\rm D}$ and $\omega_{\rm a}=\mu_{\rm a} c$, the latter
signifying the characteristic frequency of the OA signal spectrum,
is detailed in the literature, see, e.g., Ref.\,\cite{Karabutov:1998}.
Note that in response to $z_{\rm D}$ and $a_0$, the frequency
$\omega_{\rm D}$ is either decreasing or increasing, 
defining the acoustic near-field (NF) and far-field (FF)
from the value of the associated dimensionless diffraction parameter
\begin{equation}
D\, = \, \omega_{\rm D}/\omega_{\rm a} \label{eq:D} 
\end{equation}
in the regimes $D<1$ and $D>1$, respectively.

Note that the OA Volterra integral Eq.\,(\ref{eq:OAVolterraInt}) not only
allows to solve the OA forward problem, i.e.\ the calculation of the excess
pressure $p(z_{\rm D},\tau)$ given $p_0(\tau)$ and $\mathcal{K}(\tau,t)$, but also
keeps the possibility to solve the inverse OA source problem, i.e.\ the
reconstruction of $p_0(\tau)$ given $p(z_{\rm D},\tau)$ and
$\mathcal{K}(\tau,t)$, as will be detailed in the section below.


\section{Numerical implementation of the OA forward and inverse solvers}
\label{sect:Numerics}

We will now elaborate on the numerical approaches we ensued in order to solve
the OA forward and inverse problems, see subsections\ \ref{sect:Numerics_fwd}
and \ref{sect:Numerics_inv}, respectively.  In doing so we also emphasize some
important implications the special case of a Gaussian transverse irradiation
source profile has on our numerical implementation.  

\subsection{Forward solution}
\label{sect:Numerics_fwd}

Regarding the solution process of the ``direct'' OA problem, i.e.\ the
calculation of $p_{\rm D}(\tau)$ for a given distribution of initial acoustic
stress $p_{\rm 0}(\tau)$, we follow two distinct approaches.  In subsection\
\ref{subsubsect:Numerics_OAPoissonInt_fwd}, we first introduce a forward solver
based on a numeric solution of the OA Poisson integral Eq.\
(\ref{eq:OAPoissonInt}), followed by a more model-tailored solver based on a
forward solution of the OA Volterra integral Eq.\ (\ref{eq:OAVolterraInt}).
For completeness, while the optoacoustic Volterra equation can be used for
both, the calculation of $p_{\rm D}(\tau)$ as well as  for the reconstruction
of $p_{\rm 0}(\tau)$, we will subsequently need the Poisson integral based
solver for benchmarking and generation of synthetic OA signals that serve as
input for the inversion procedure.  This is necessary in order to formally
de-couple the forward solution and inversion processes.

\subsubsection{Forward solver based on the OA Poisson integral:}
\label{subsubsect:Numerics_OAPoissonInt_fwd}

Here, we opt for an implementation of Eq.\ (\ref{eq:OAPoissonInt}) for layered
media in cylindrical polar coordinates. The respective implementation is
available as ``SONOS -- a fast poisson integral solver for layered homogeneous
media'' \cite{comment:GitHub_OACode1}, and, albeit being restricted to a solely
$z$-dependent absorption coefficient, it allows for the efficient calculation
of OA signals resulting from general irradiation source profiles with an axial
symmetry at arbitrary detection points~$\vec{r}_{\rm D}=(\rho_{\rm D},z_{\rm
D})$.  Therein, $z_{\rm D}$ signifies the axial coordinate of the detection
point in the reference frame in which the nearest absorbing layer has $z=0$,
see Fig.\ \ref{fig:setup}, and $\rho_{\rm D}$ denotes the deviation of the
detection point from the symmetry axis of the beam profile.  Here, since we are
only interested in the calculation of OA signals along the beam axis ($\rho_{\rm D}=0$), we can further simplify the numerical procedure detailed in
Ref.~\cite{Blumenroether:2016} to some extend. Since the on-axis view of
the irradiation source profile in the detection-point based reference frame is
independent of the azimuthal angle, and therefore $f_{\rm D}(\rho)=f_0
\exp\{-\rho^2/a_0^2\}$, the respective integration in a cylindrical polar
representation of Eq.\ (\ref{eq:OAPoissonInt}) can be carried out explicitly,
resulting in the simplified expression
\begin{equation}
p_{\rm D}(t) = \frac{\Gamma}{2 c} \partial_t
\int\limits_{\rho}\!\int\limits_{z}\!\rho\frac{f_{\rm D}(\rho) g_{\rm D}(z)}{(\rho^2 +
z^2)^{1/2}} \delta((\rho^2 + z^2)^{1/2} -
ct)\,\mathrm{d}\rho\,\mathrm{d}z. \label{eq:OAPoissonInt_D}
\end{equation}
Note that in the above equation, the distance $z$ is to be measured with
respect to the detection point ${\rm D}$, with the nearest absorbing layer
located at $z=|z_{\rm D}|$. Further, the $\delta$-distribution might be
interpreted as an indicator function that bins the values of the integrand
according to the propagation time of the respective stress waves.  This already
yields a quite efficient numerical scheme to compute the OA signal $p_{\rm
D}(t)$ at the detection point, since the pending integrations can, in a
discretized setting where 
$\rho_i=i\Delta_\rho$,
$i=0,\ldots,N_{\rho}$, and 
$\Delta_\rho=\rho_{\rm max}/N_{\rho}$ as well as
$z_i=|z_{\rm D}|+i\Delta_z$, 
$i=0,\ldots,N_{z}$, and 
$\Delta_z=(z_{\rm max} - |z_{\rm D}| )/N_{z}$, 
be carried out with time complexity of order $O(N_\rho N_z)$.
During our numerical experiments, since we are only interested in the general
shape of the optoacoustic signal, we set the value of the constants in Eq.\
(\ref{eq:OAPoissonInt_D}) to $\Gamma/c \equiv 2/f_0$.  Thus, the resulting
OA signal is obtained in arbitrary units, subsequently abbreviated
as $[{\rm a.u.}]$. Finally, so as to mimic the finite thickness $\Delta w$ of
the transducer foil in an experimental setup \cite{Blumenroether:2016}, we
grant the option to average $p_{\rm D}(t)$ at the detection point over a time 
interval $\Delta t = \Delta w/c$.

\subsubsection{Forward solver based on the OA Volterra integral:}
\label{subsubsect:Numerics_OAVolterraInt_fwd}

While there exist standard procedures for the numerical (forward) solution of
Volterra integral equations of the $2$nd kind, e.g.\ based on an approximation
of the diffraction term in Eq.\ (\ref{eq:OAVolterraInt}) in terms of a
trapezoidal rule \cite{NumRec:1992} (or other quadrature rules
\cite{Hazewinkel:1987}, for that matter), we can simplify the approach for a 
general kernel $\mathcal{K}(\tau,t)$ by capitalizing on the special 
form of the OA stress wave propagator. Since the latter is of convolution
type, i.e.\ $\mathcal{K}(\tau,t)\equiv\mathcal{K}(\tau-t)=
\omega_{\rm D}\exp\{-\omega_{\rm D} (\tau-t)\}$, Eq.\ (\ref{eq:OAVolterraInt}) 
can be solved for $p_{\rm D}(\tau)$
via memoization \cite{Michie:1968}. Therefore, and with regard to the upcoming
inversion step in subsection\ \ref{subsubsect:Numerics_OAVolterraInt_inv}, it is
beneficial to put the diffraction term in Eq.\ (\ref{eq:OAVolterraInt}) under
scrutiny. As it turns out, in a discretized setting where 
$t_i=i\Delta_t$, $i=0,\ldots,N$, $\Delta_t=t_{\rm max}/N$, and thus
$\tau_i = t_i+z_{\rm D}/c$,
abbreviating $\mathcal{K}(\tau_i-\tau_j)$ and $p_{\rm 0}(\tau_i)$ as $\mathcal{K}_{i,j}$ and
$p_{{\rm 0},i}$, respectively, we have
$\mathcal{K}_{i,i}=\omega_{\rm D}$ and $\mathcal{K}_{i+2,i}=
\mathcal{K}_{i+1,i} \exp\{-\omega_{\rm D} \Delta_t\}$, 
yielding a recurrence relation that approximates the diffraction term according
to 
\begin{eqnarray}
I_{i} &= \int_{\tau_0}^{\tau_{i}}\!\mathcal{K}(\tau_{i}-t^\prime) p_{\rm 0}(t^\prime)\,\mathrm{d}t^\prime \nonumber\\ 
&= \Big[ \mathcal{K}_{i,0}p_{{\rm 0},0} + 2 \sum_{j=1}^{i-1}\mathcal{K}_{i,j} p_{{\rm 0},j} + \mathcal{K}_{i,i} p_{{\rm 0},i} \Big]\,\Delta_t/2 \nonumber\\
&= I_{i-1} \exp\{-\omega_{\rm D} \Delta_t\} + (\omega_{\rm D} \Delta_t/2) \Big[ p_{{\rm 0},i-1} \exp\{-\omega_{\rm D}\Delta_t\} + p_{{\rm 0},i} \Big]. \label{eq:OAVolterra_recRel}
\end{eqnarray}
In the above expression the trapezoidal approximation becomes exact in the
limit $N \to \infty$.  Consequently, the OA signal $p_{{\rm D},i}\equiv
p_{\rm D}(\tau_i)$ can be obtained by simply marching in time, i.e.\ 
\begin{equation}
p_{{\rm D},i} = p_{{\rm 0},i} - I_i, \qquad (i=1,\ldots,N) \label{eq:OAVolterra_fwdSolver}
\end{equation}
starting off at $p_{{\rm D},0}=p_{{\rm 0},0}$, $I_0=0$ and updating $I_i$ via 
Eq.\ (\ref{eq:OAVolterra_recRel}).

Note that adopting a standard discretized scheme for the calculation of
$p_{\rm D}(\tau)$, set up for a general kernel $\mathcal{K}(\tau,t)$
\cite{NumRec:1992}, would yield an algorithm that terminates in time $O(N^2)$ 
since the full integral has to be re-computed at each time-step due to the 
``wandering'' upper bound. Note that this would not mean much an improvement
over the full wave equation solver discussed in subsection 
\ref{subsubsect:Numerics_OAPoissonInt_fwd} as one has $N\approx N_z$. 
However, since the OA stress wave propagator is
actually of convolution type and can be decomposed into a product of exponential
factors at each time-step, we here yield a highly efficient custom forward 
solver with time complexity $O(N)$.

\subsection{Inverse solution}
\label{sect:Numerics_inv}

We can accomplish the highly non-trivial task of solving the inverse OA
source problem, i.e.\ the reconstruction of $p_{{\rm 0}}(\tau)$ given $p_{\rm
D}(\tau)$ and $\mathcal{K}(\tau,t)$ via Eq.~(\ref{eq:OAVolterraInt}), in
basically two ways: by means of (i) an inversion scheme that complements
the Volterra integral based forward solver presented earlier, see
subsection\ \ref{subsubsect:Numerics_OAVolterraInt_inv}, and, (ii) an independent
algorithmic procedure based on the idea of iteratively improving a putative
initial solution, see subsection\ \ref{subsubsect:Numerics_PicardIter_inv}.

\subsubsection{Inverse solver based on the OA Volterra integral:}
\label{subsubsect:Numerics_OAVolterraInt_inv}
In terms of the recurrence relation approach that yields the diffraction term
in Eq.~(\ref{eq:OAVolterraInt}) at time-step $i$ via memoization, see
Eq.~(\ref{eq:OAVolterra_recRel}), the actual inversion step is only slightly
more involved than the forward solution. I.e., the inverse solution in terms of
the OA Volterra integral equation can be accomplished by updating the values of
$p_{{\rm D},i}$ and $I_i$ in a leap-frog manner according to
\begin{eqnarray}
p_{{\rm 0},i} &= \Big(1-\omega_{\rm D} \Delta_t/2 \Big)^{-1} 
\Big[  p_{{\rm D},i} + \Big( I_{i-1} + (\omega_{\rm D}\Delta_t/2) p_{{\rm 0},i-1} \Big) \exp\{-\omega_{\rm D}\Delta_t\} \Big] \\
I_i &= I_{i-1} \exp\{-\omega_{\rm D} \Delta_t\} + (\omega_{\rm D} \Delta_t/2) \Big[ p_{{\rm 0},i-1} \exp\{-\omega_{\rm D}\Delta_t\} + p_{{\rm 0},i} \Big], \nonumber
\end{eqnarray}
starting off at $p_{{\rm 0},0}=p_{{\rm D},0}$ and $I_0=0$. Thus, the numerical
expense of the Volterra integral based inverse solver amounts to only $O(N)$.

\paragraph{Far-field inversion --}
Considering a far-field (FF) setup wherein the distance between the detection
point ${\rm D}$ and the absorbing layers is large, i.e.\ $z_{\rm D}\to \infty$
and the width of the irradiation source profile is narrow enough to ensure a
diffraction parameter $D\gg 1$ (also referred to as Fraunhofer zone
\cite{Sigrist:1986}; a parameter region also important for approximate OA
imaging methods \cite{Burgholzer:2007,Mitsuhashi:2014}), the OA signal $p_{{\rm
D},{\rm FF}}(\tau)$ is related to the initial OA stress profile $p_{\rm
0}(\tau)$ via \cite{Karabutov:1998,Paltauf:2000}
\begin{equation}
p_{{\rm D},{\rm FF}}(\tau)= \frac{1}{\omega_{\rm D}} \frac{\mathrm{d}}{\mathrm{d}\tau}p_{\rm 0}(\tau). \label{eq:inversion_FF}
\end{equation}
Thus, in the far-field approximation the initial stress profile $p_{{\rm
0},{\rm FF}}(\tau)$  can be obtained by numerical quadrature using the above
equation.

\subsubsection{Inverse solver based on successive approximations:}
\label{subsubsect:Numerics_PicardIter_inv}

For numerical redundancy and so as to establish a broad computational
foundation for the Volterra integral based approach to inverse optoacoustics,
we employed a further, independent reconstruction scheme.  It relies on the
continued refinement of a putative solution in terms of the Picard-Lindel\"of
iteration method \cite{Hairer:1993}, wherein a properly guessed first
approximation $p_{\rm 0}^{(0)}(\tau)$ of $p_{\rm 0}(\tau)$ is improved
successively by solving 
\begin{equation}
p_{\rm 0}^{(n+1)}(\tau) = p_{\rm D}(\tau) + \int_{-\infty}^\tau \!\mathcal{K}(\tau-t)\, p_{\rm 0}^{(n)}(t)\,\mathrm{d}t.
\end{equation}
I.e.\ the overall numerical effort to advance from approximation $n\to n+1$ 
in the above correction procedure  amounts to a numerical integration that 
can be accomplished in a time linear in the number of interpolation steps.  
On very general grounds, the Chebyshev-norm of the difference between two 
successive solutions \mbox{$c_n \equiv ||p_{\rm}^{(n+1)}(\tau)-p_{\rm 0}^{(n)}(\tau)||$} becomes
arbitrarily small as $n \to \infty$ and $p_{\rm 0}(\tau)~=~\lim_{n\to\infty}
p_{\rm 0}^{(n)}(\tau)$. 
From a practical point of view we terminated the iteration scheme as soon as
the above norm decreases below the threshold $c_n \leq
10^{-6}$, observing that the procedure converges quite fast, i.e.\ within
$O(10-60)$ approximation cycles, see the discussion in section\
\ref{sect:NumEx_picardIt}.
Note that thus, the time complexity of the solver is basically limited by the
time spent for the repeated computation of the integrating term.  For a proper
choice of a ``predictor'' $p_{\rm 0}^{(0)}(\tau)$ one might distinguish two main
reconstruction regimes:
\paragraph{Near-field reconstruction --}
As a high-precision predictor we here use the initial guess $p_{\rm
0}^{(0)}(\tau)\equiv p(z_{\rm D},\tau)$ since we can expect the OA near-field signal
to be still quite close to the distribution of initial stress $p_{\rm 0}(\tau)$.
In contrast to this, a low-precision predictor can be obtained by setting
$p_{\rm 0}^{(0)}(\tau) \equiv p_{\rm 0}$, where e.g.\ $p_{\rm 0}=1$.

\paragraph{Far-field reconstruction --}
As a high-precision predictor we might use the initial guess 
$p_{\rm 0}^{(0)}(\tau)\equiv p_{{\rm 0},{\rm FF}}(\tau)$, obtained by integrating
the OA signal $p(z_{\rm D},\tau)$ in the far-field approximation 
Eq.\ (\ref{eq:inversion_FF}).
In opposition to this, a low-precision predictor can be obtained by again
setting $p_{\rm 0}^{(0)}(\tau) \equiv p_{\rm 0}$.



\begin{figure}[t!]
\centering
  \begin{minipage}[b]{75 mm}
    \centering\includegraphics[trim = 30mm 145mm 15mm 42mm, clip, width=8cm]{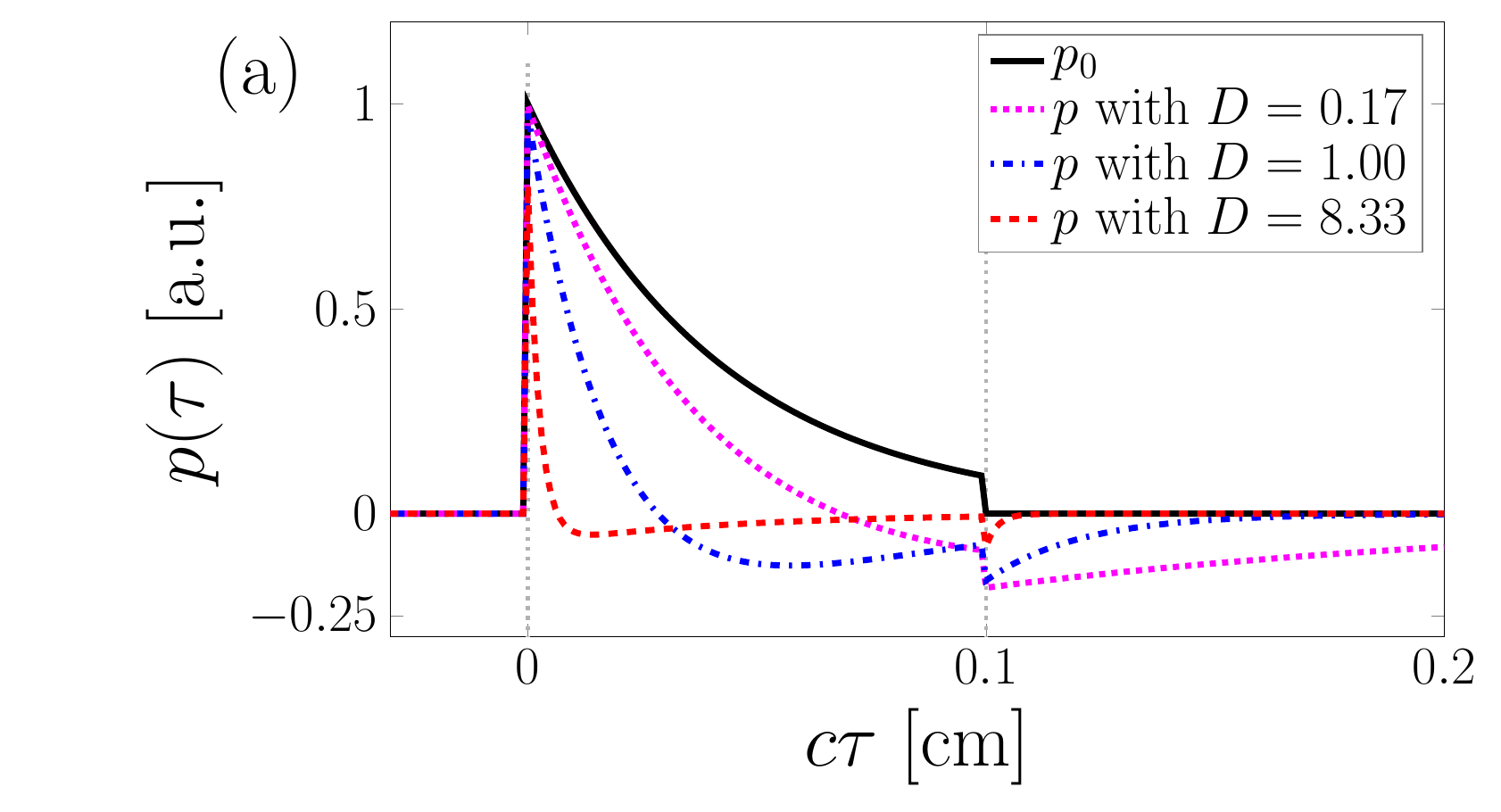}  
  \end{minipage}
  \begin{minipage}[b]{75 mm}
    \centering\includegraphics[trim = 30mm 145mm 15mm 42mm, clip, width=8cm]{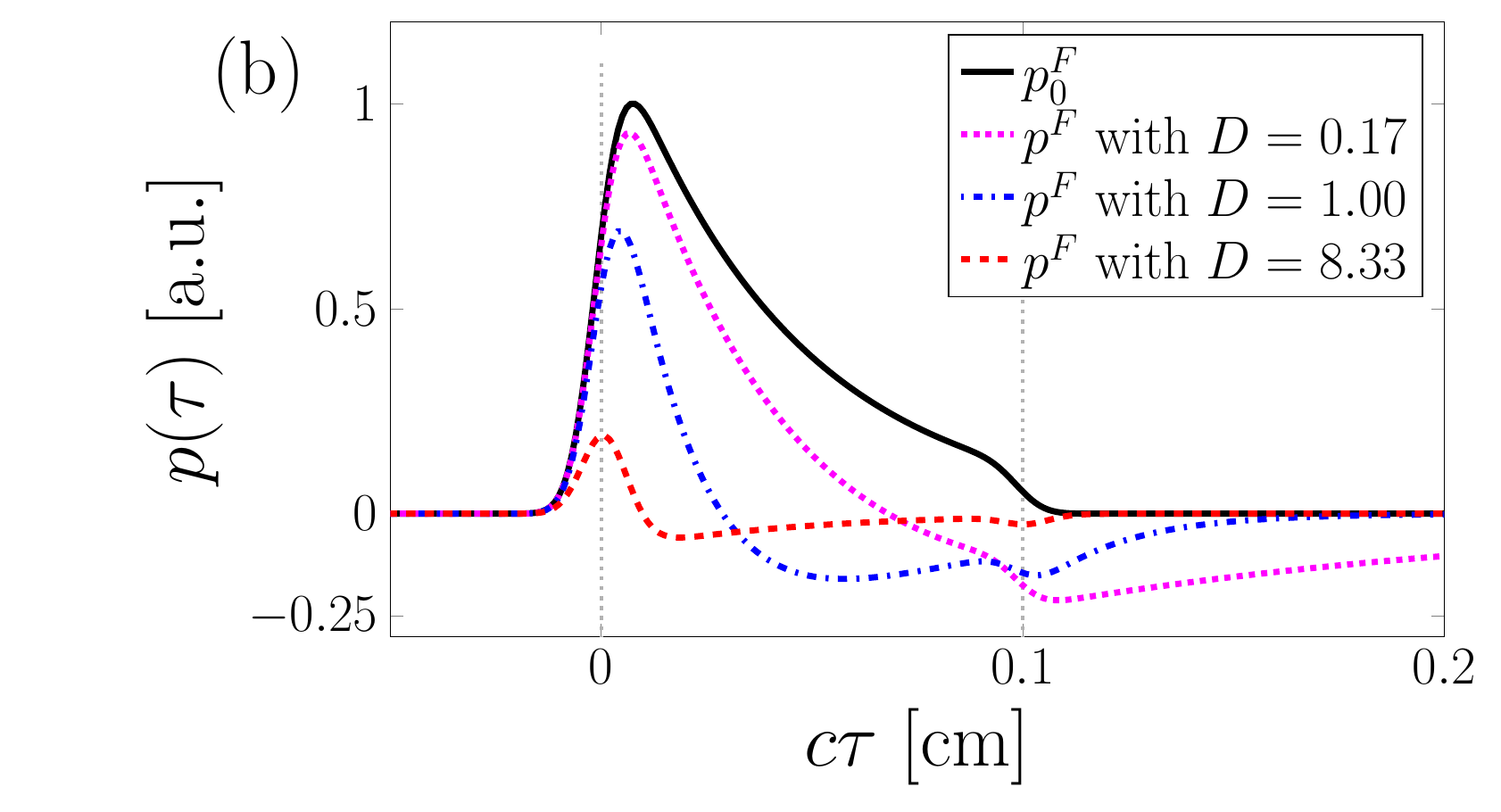} 
  \end{minipage}
\caption{(Color online) Forward calculation of optoacoustic signals $p(\tau)$ in
the framework of the Volterra integral equation.  In either subfigure, $p_{\rm
0}$ (black solid lines) indicates the on-axis profile of the initial stress
according to Eqs.~(\ref{eq:iniStress}-\ref{eq:isp}). The figures illustrate the
change in shape of the optoacoustic signals as perceived at detection points
with increasing detector-to-layer distance $z_{\rm D}=-0.02~{\rm cm}$ (magenta
dotted line), $-0.1~{\rm cm}$ (blue dash-dotted line), $-1~{\rm cm}$ (red
dashed line), characterized by the diffraction parameters $D=0.17$, $1.0$, and,
$8.33$, respectively. 
(a)~forward solution of a ``genuine'', i.e.\ non-preprocessed initial stress
profile $p_{\rm 0}$, 
(b)~forward solution of an $\Delta_z=0.01~{\rm cm}$ sliding average 
Gaussian smoothed initial stress profile.
}
\label{fig:Volterra_direct}
\end{figure}

\section{Numerical experiments} 
\label{sect:NumEx}

The simulation parameters used for the subsequent numerical experiments are
geard towards actual ``laboratory'' parameters for existing polyvinyl alcohol
hydrogel (PVA-H) based tissue phantoms  
used in the combined experimental and numerical study reported in Ref.\
\cite{Blumenroether:2016}. These consist of melanin doped absorbing layers in
between two layers of clear PVA-H, quite similar to the setup illustrated in
Fig.\ \ref{fig:setup}.  Here, we assume that the PVA-H layers are
non-absorbing and that the melanin doped layer exhibits an absorption coefficient of
$\mu_a = 24~{\rm cm^{-1}}$. Albeit the irradiation source profile
in such an experimental setting is usually assumed to be of ``top-hat'' type
\cite{Paltauf:2000,Paltauf:1998,Blumenroether:2016} (e.g.\ Ref.\
\cite{Blumenroether:2016} report an overall $1/e$ radius of approximately $1.2~{\rm
cm}$), we here assume an effective Gaussian beam profile with $1/e$ radius
$a_0=0.1~{\rm cm}$.  This stimulating light beam meets the absorbing layer at
$z = 0$ and leaves it at $z = \Delta z$. The resultant acoustic signal is
computed for a field point located along the beam axis at position 
$z=z_{\rm D}<0$. 

\subsection{Forward and inverse solution within the OA Volterra framework}

In the first series of numerical experiments, we deliberately stayed within the
framework of the paraxial approximation \cite{comment:inverseCrime}, i.e.\ we
accomplished the forward and inverse calculations by means of the solvers
derived from the optoacoustic Volterra equation. Such an approach
might be considered as committing inverse crime \cite{Colton:2012}, i.e.\
performing a (putative) trivial inversion of synthetic data obtained by first
solving the forward problem in terms of the same exact model. However, here we
use this approach as a proof of principle and consider an independent forward
solver (with no connection to the Volterra based solver) in subsequent
sections. 
Albeit there is a wealth of literature discussing OA signals and their change in
shape upon advancing from a measurement point located in the acoustic
near-field (NF) to a point in the far-field (FF), see e.g.\ Refs.\
\cite{Gusev:1993,Karabutov:1998,Sigrist:1986}, we first aimed at briefly 
illustrating the respective diffraction triggered signal changes for our 
particular single-layer setup by considering different values of $z_{\rm D}$.  

\begin{figure}[t!]
\centering
  \begin{minipage}[b]{75 mm}
    \centering\includegraphics[trim = 30mm 148mm 20mm 42mm, clip, width=8cm]{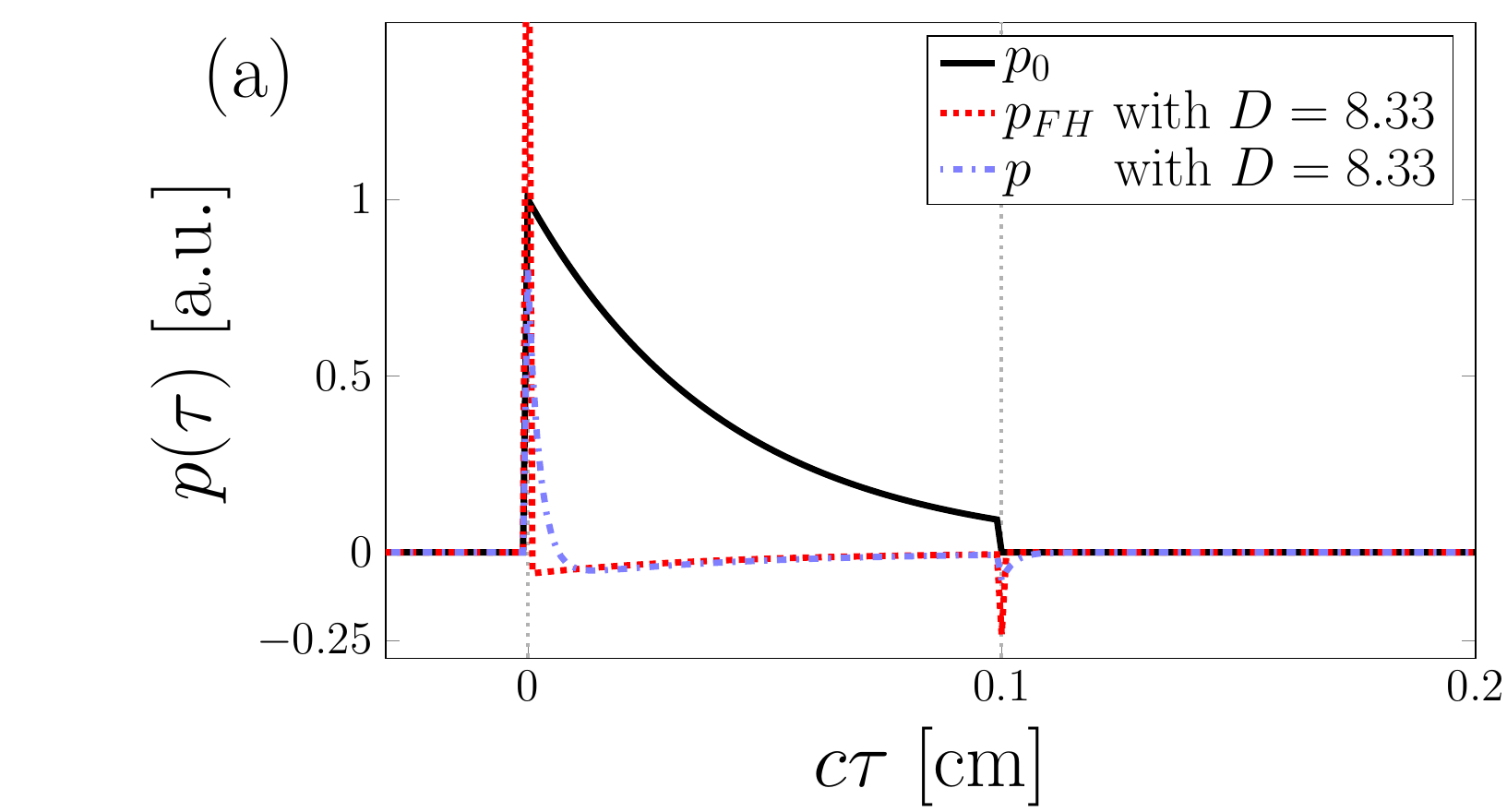}  
  \end{minipage}
  \begin{minipage}[b]{75 mm}
    \centering\includegraphics[trim = 30mm 148mm 20mm 42mm, clip, width=8cm]{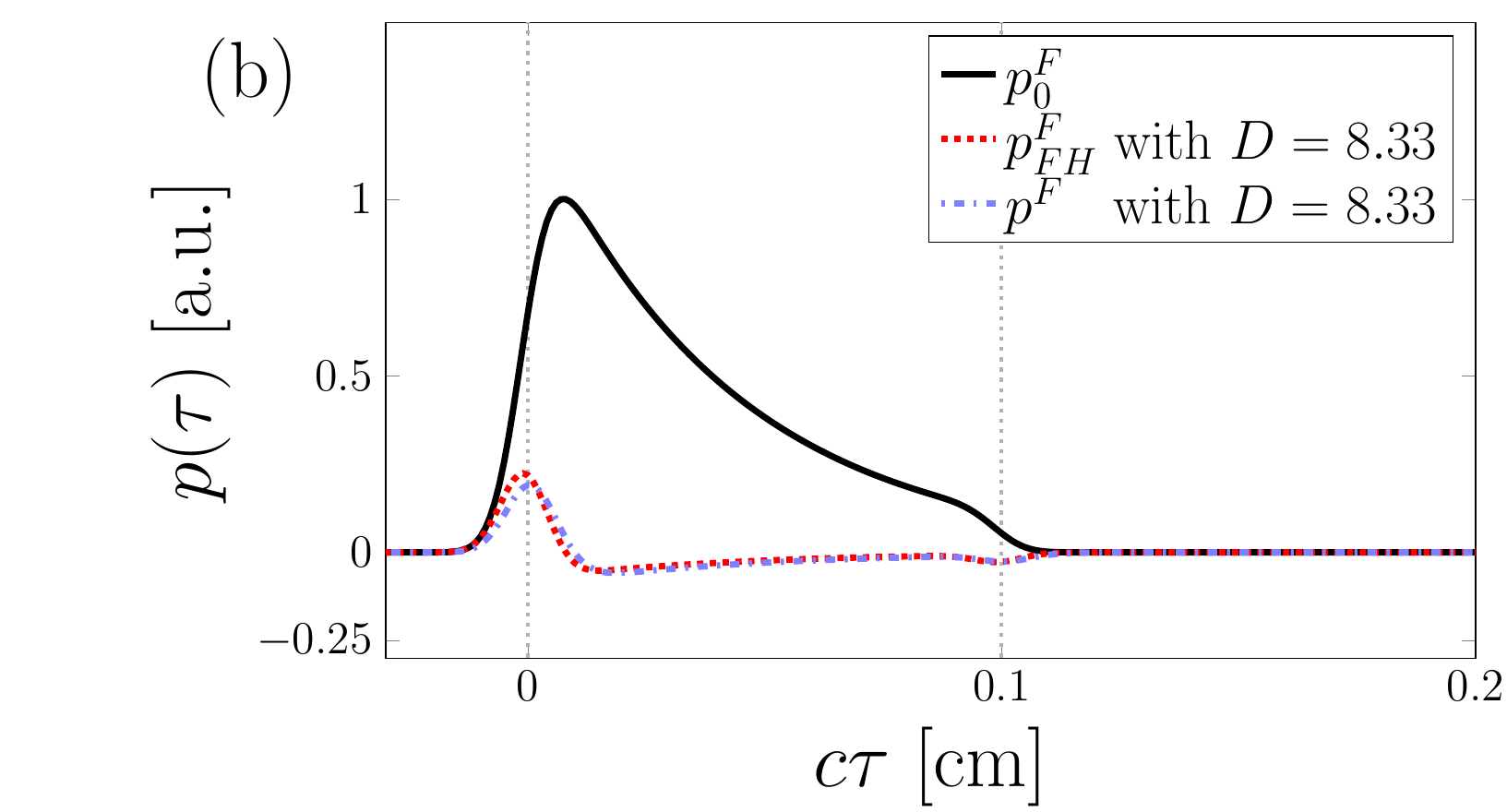} 
  \end{minipage}
\caption{(Color online)
Forward calculation of optoacoustic signals in
a far-field approximation considering Eq.\ (\ref{eq:inversion_FF}).  In either
subfigure, $p_{\rm 0}$ (black solid lines) indicates the on-axis profile of the
initial acoustic stress. The figures illustrate the difference in the
approximate FF formula, resulting in the signal $p_{\rm FH}$ (red dashed curve),
relative to the exact solution $p$ (red dash-dotted curve) at $z_{\rm
D}=-1~{\rm cm}$, i.e.\ $D=8.33$.  (a) forward solution of a ``genuine'', i.e.\
non-preprocessed initial stress profile $p_{\rm 0}$, (b) forward solution of a
smoothed initial stress profile ($\Delta_z=0.01~{\rm cm}$ sliding average
Gaussian smooth).}
\label{fig:Fraunhofer_direct}
\end{figure}

\paragraph{Solving the direct problem --}
In this regard, Fig.\ \ref{fig:Volterra_direct} illustrates the forward
calculation of OA signals in a discretized setting, where $z_i=z_{\rm min}+i
\Delta_z$, $i=1,\ldots, N$, and $\Delta_z=(z_{\rm max}-z_{\rm min})/N$,
starting from an absorption profile obtained using Eqs.\ (\ref{eq:iniStress})
and (\ref{eq:g}), i.e.\ following Beer-Lamberts law for pure absorbers.  Here,
we used $z_{\rm min}=0.0~{\rm cm}$, $z_{\rm max}=0.1~{\rm cm}$, $N=300$ as well
as $c=1$, thus $\tau_i=z_i$.  While Fig.\ \ref{fig:Volterra_direct}(a) relates to
the forward solution of the OA problem for a non-preprocessed distribution of
initial stress, Fig.\ \ref{fig:Volterra_direct}(b) refers to a smoothed initial
condition where $p_{\rm 0}$ is preprocessed using a $\Delta_z=0.01~{\rm cm}$
sliding average Gaussian filter to mimic a laboratory scenario wherein the
increase and decrease in absorption coefficient is less sudden. 
Note that the general shape of the OA NF signal, characterized by the
diffraction parameter $D=0.17$, is still strongly reminiscent of the shape of
the absorption profile. However, the initial compression phase has already
noticeably transformed by diffraction, giving rise to an extended rarefaction
phase above a retarded signal depth of $\approx 0.08~{\rm cm}$, extending well
beyond the signal depth that characterized the end of the absorbing layer.  In
contrast to this, the borderline FF ($D=1.0$) and FF ($D=8.33$) signals allow
for a proper OA depth profiling: both feature a pronounced compression peak
that signals an increase of the absorption coefficient $\mu_a$ at $z=0~{\rm
cm}$, followed by an extended rarefaction phase until a sharp (in case of Fig.\
\ref{fig:Volterra_direct}(a); smooth cusp in case of Fig.\
\ref{fig:Volterra_direct}(b)) rarefaction dip signals a sudden decrease of the
absorption coefficient at $z=0.1~{\rm cm}$ continued by a further rarefaction
phase rapidly decaying in amplitude.  
Note that Fig.\,\ref{fig:Volterra_direct}(b) boldly reveals a particular
property of OA FF signals: since in the acoustic FF $p(\tau)$ is related to
$p_{\rm 0}(\tau)$ via differentiation, see Eq.\ (\ref{eq:inversion_FF}), the
peak value of the initial compression phase shifts towards the inclination
point of the leading edge of $p_{\rm 0}(\tau)$ as $D$ increases.
As evident from Fig.\ \ref{fig:Fraunhofer_direct}, the simplified calculation
of the OA signal in the FF approximation according to Eq.\
(\ref{eq:inversion_FF}) is already quite precise at $D=8.33$, improving further
in quality as $|z_{\rm D}|\to \infty$ (not shown). Albeit there is still a
slight difference between the FF signal estimator $p_{\rm FF}(\tau)$ and the exact
signal shape $p(\tau)$ for the genuine distribution of initial stress (see Fig.\
\ref{fig:Fraunhofer_direct}(a)), the difference seems less pronounced in case
of the smoothed initial stress configuration shown in Fig.\
\ref{fig:Fraunhofer_direct}(b).

\begin{figure}[t!]
\centering
  \begin{minipage}[b]{75 mm}
    \centering\includegraphics[trim = 30mm 145mm 15mm 42mm, clip, width=8cm]{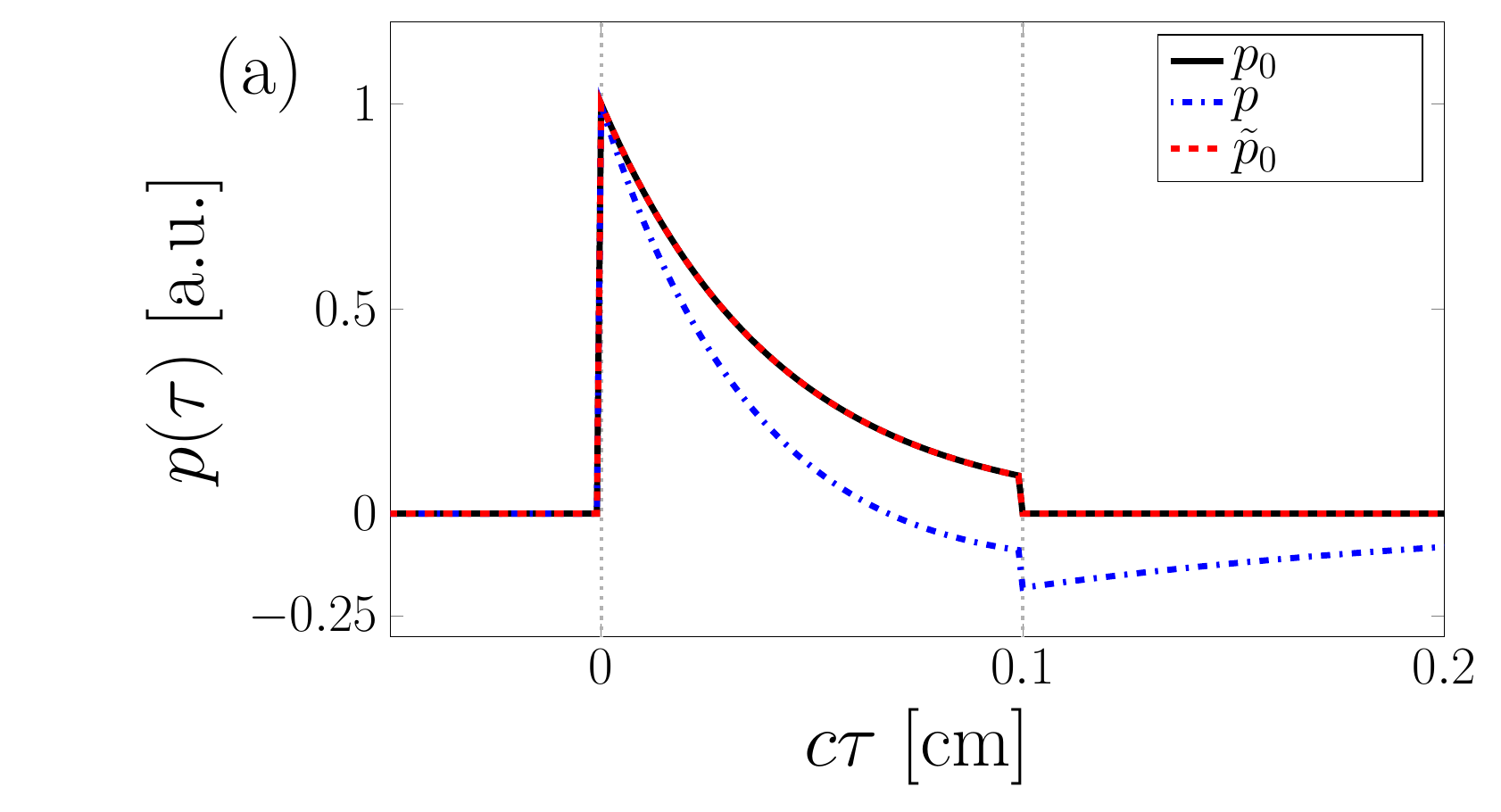}  
  \end{minipage}
  \begin{minipage}[b]{75 mm}
    \centering\includegraphics[trim = 30mm 145mm 15mm 42mm, clip, width=8cm]{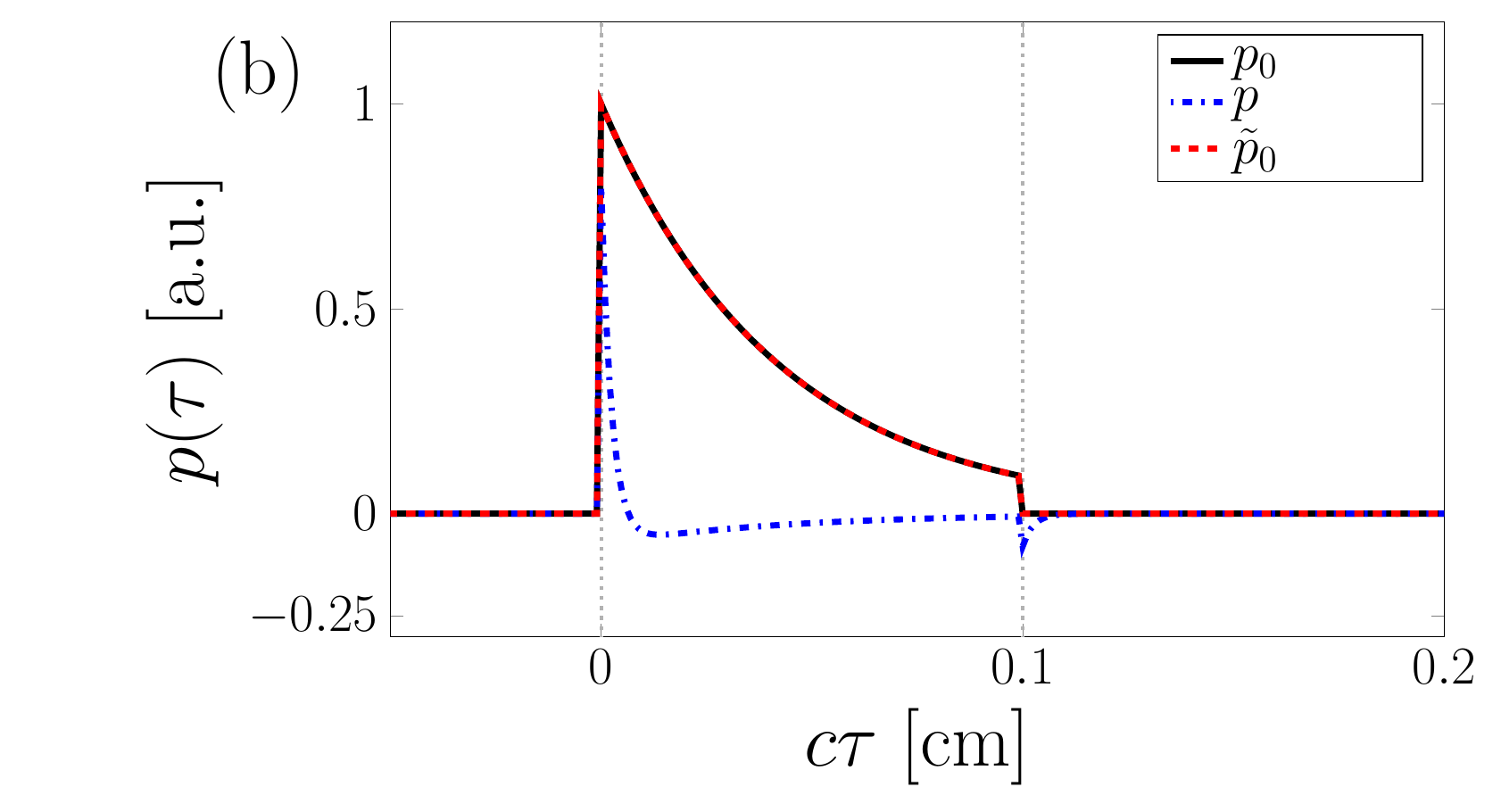} 
  \end{minipage}
\caption{(Color online) 
Reconstruction of the initial acoustic stress $\tilde{p}_{\rm 0}$ from computed 
OA signals $p$ upon knowledge of the diffraction propagator $\mathcal{K}$ in 
(a) the acoustic NF at $z_{\rm D}=-0.02~{\rm cm}$, i.e.\ $D=0.17$, and
(b) the acoustic FF at $z_{\rm D}=-1.0~{\rm cm}$, i.e.\ $D=8.33$.
The OA source reconstruction is accomplished with the leap-frog algorithm 
introduced in subsect.\ \ref{subsubsect:Numerics_OAVolterraInt_inv}.
In either case, the reconstructed stress profiles perfectly match the true
initial stress profiles $p_{\rm 0}$.}
\label{fig:Inverse}
\end{figure}

\paragraph{Solving the inverse OA source reconstruction problem --}

In a second set of numerical experiments we aimed at solving the inverse
optoacoustic problem, where the aim is to reconstruct the initial distribution
of acoustic stress $p_0$ from the measured signal $p$ upon knowledge of
the optoacoustic stress propagator $\mathcal{K}$.  As evident from Fig.\
\ref{fig:Inverse}, the Volterra Integral equation based inverse solver outlined
in subsection\ \ref{subsubsect:Numerics_OAVolterraInt_inv} accomplishes this task
in efficient fashion: irrespective of whether the inversion is performed in the
acoustic NF or FF, see Figs.\ \ref{fig:Inverse}(a) and (b), respectively, the
reconstructed stress profiles $\tilde{p}_{\rm 0}$ perfectly match the exact
initial stress profiles $p_{\rm 0}$.

\subsection{Poisson based forward and Volterra inverse solution}
\begin{figure}[t!]
\centering
  \begin{minipage}[b]{75 mm}
    \centering\includegraphics[trim = 30mm 145mm 15mm 42mm, clip, width=8cm]{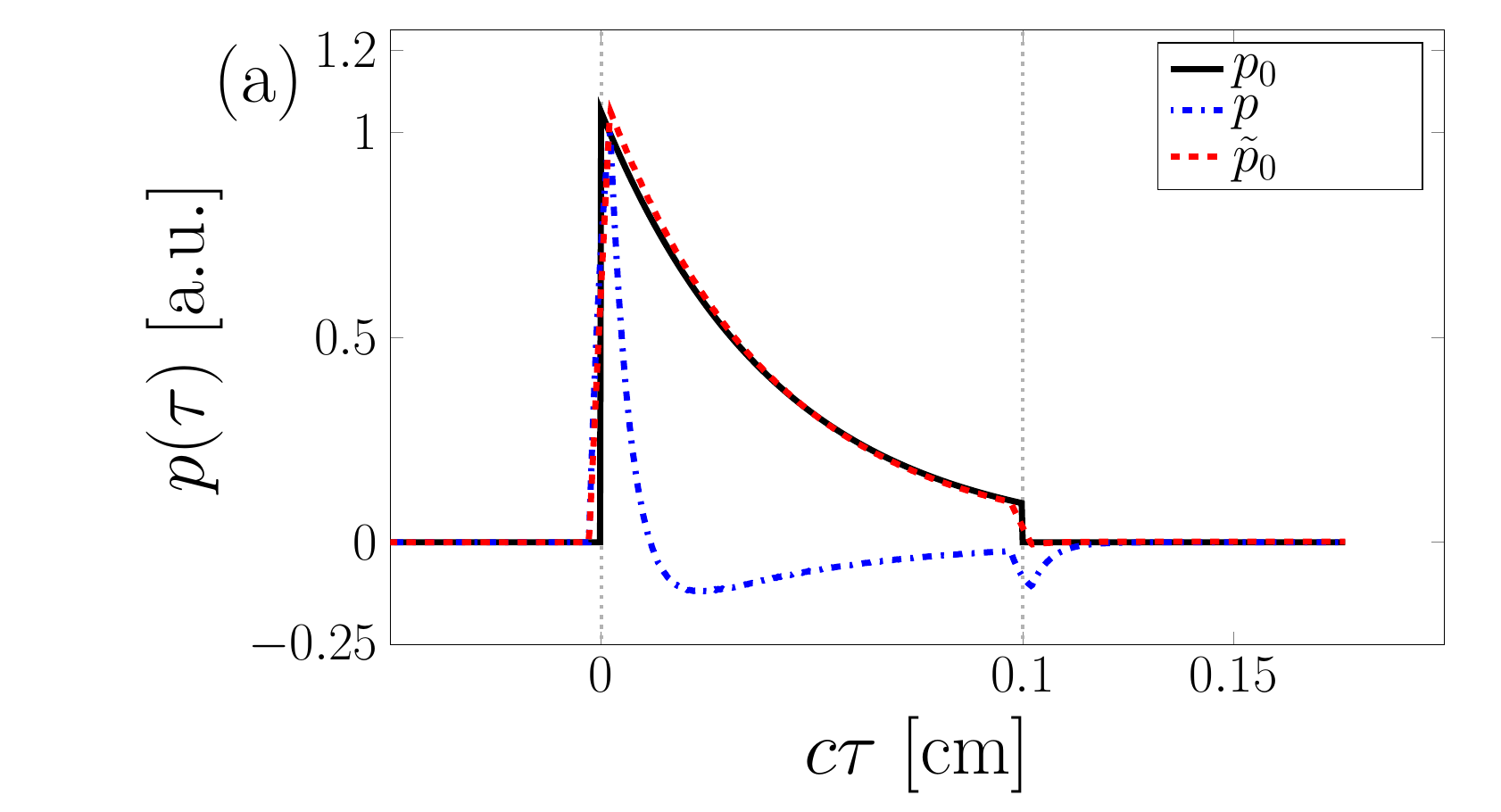}  
  \end{minipage}
  \begin{minipage}[b]{75 mm}
    \centering\includegraphics[trim = 30mm 145mm 15mm 42mm, clip, width=8cm]{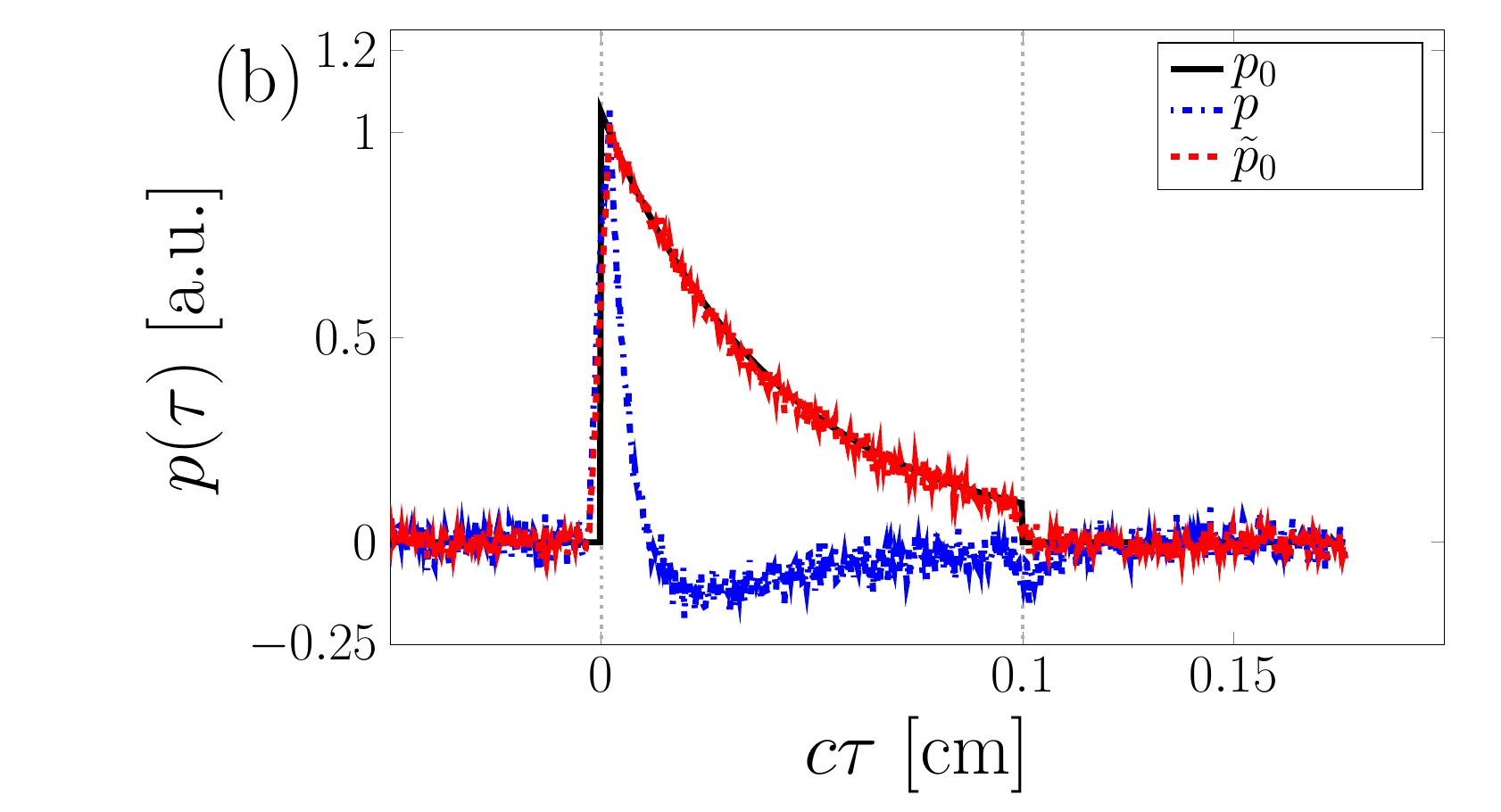} 
  \end{minipage}
\caption{
(Color online) Solution of the source reconstruction problem using the Volterra
integral equation based inverse solver. The figure illustrates the
reconstruction of the initial acoustic stress profile $\tilde{p}_{\rm
0}$ (dashed red line) from a far-field signal $p$ (dash-dotted blue
line), calculated at $z_{\rm D}=-1~{\rm cm}$.
The true initial stress profile is given by $p_{\rm 0}$ (solid black line).
(a) reconstruction starting from a genuine, i.e.\ non-postprocessed OA signal
$p$.
(b) reconstruction in presence of noise. To obtain the noisy signal, the 
genuine OA signal was superimposed by Gaussian white-noise with a signal-to-noise
ratio of $5$.
}
\label{fig:Volterra_inverse}
\end{figure}

While the above computation and inversion of OA signals were performed using
solvers, both derived within the framework of the OA Volterra integral
equation, we subsequently consider the independent forward solver ``SONOS''
\cite{comment:GitHub_OACode1}, detailed in subsection\
\ref{subsubsect:Numerics_OAPoissonInt_fwd}.  It yields synthetic input data
based on the solution of the full wave equation and, bearing in mind that the
inversion is accomplished in the paraxial approximation, consequently helps to
prevent inverse crime.  Note that, since the paraxial approximation on which
the OA Volterra integral equation is based describes the underlying wave
equation best at sufficiently large distances $|z_{\rm D}|$, we first focus on
the inversion of signals in the acoustic FF.  An exemplary inversion procedure
considering the Poisson based forward solver and Volterra based inverse solver
at $z_{\rm D}=-1.0~{\rm cm}$ ($D=8.33$) is shown in
Fig.~\ref{fig:Volterra_inverse}. As evident from the figure, the initial
acoustic stress profile $p_{\rm 0}$ features sharp edges at the boundary of the
absorbing layer whereas the inverse estimate $\tilde{p}_{\rm 0}$ is more gently
inclined. This is due to the temporal average over a time interval $\Delta t =
0.005~{\rm cm}/c$ which is used to mimic the finite extension of a detector in
an experimental setting (see discussion in section\
\ref{subsubsect:Numerics_OAPoissonInt_fwd} and Ref.\
\cite{Blumenroether:2016}).

\begin{figure}[t!]
\centering\includegraphics[trim = 30mm 138mm 15mm 42mm, clip, width=8cm]{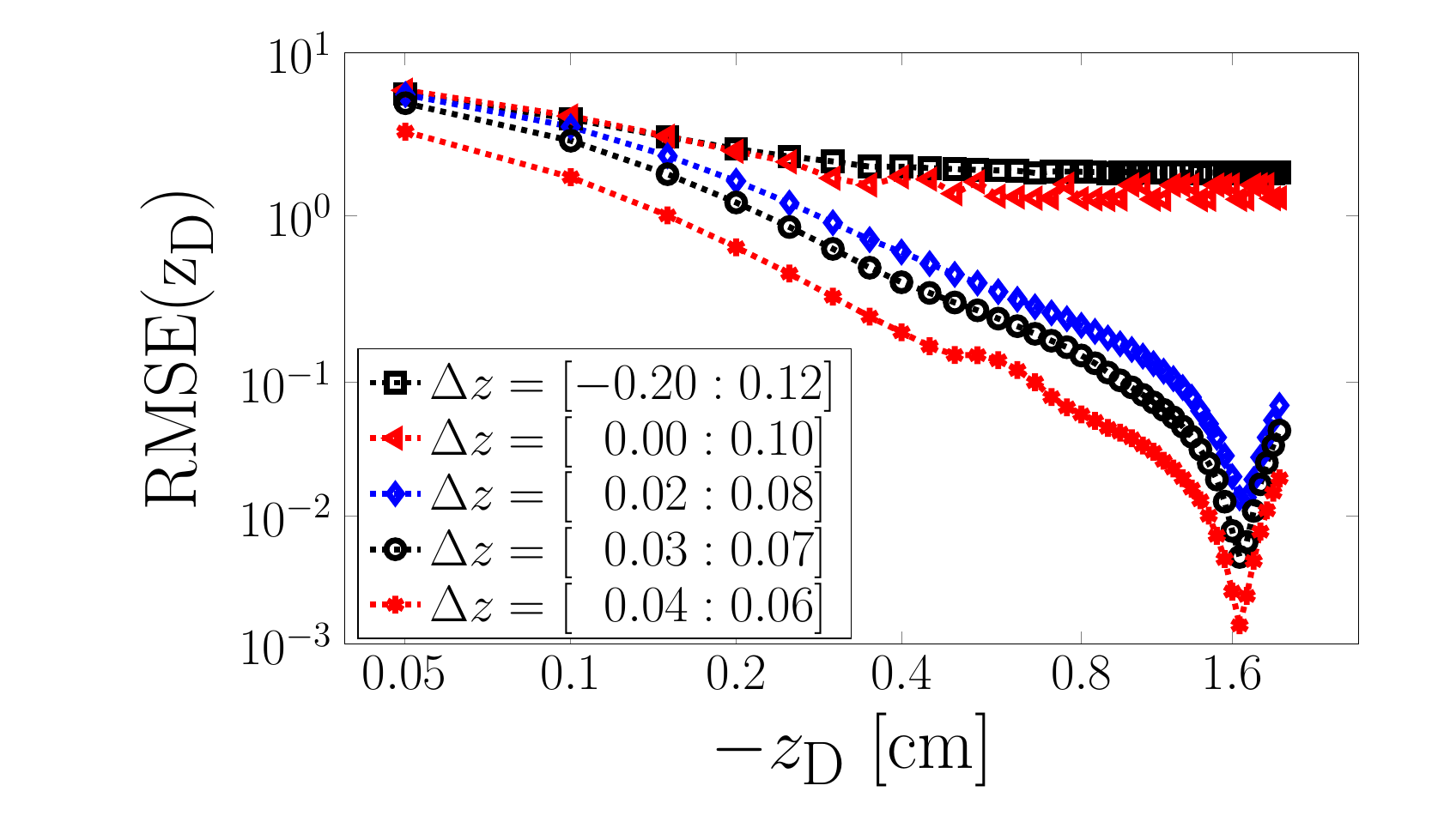}  
\caption{
Root mean squared error (RMSE) between the true initial stress profile 
$p_{\rm 0}$ and
the inverse estimate $\tilde{p}_{\rm 0}$ for increasingly narrow intervals
$\Delta z \equiv [c \tau_{-}:c \tau_{+}]$ along the $z$- and $\tau$-axis,
respectively. Note that, in a preprocessing step, $p_{\rm 0}$ was 
normalized and $\tilde{p}_{\rm 0}$ amplitude-adjusted as detailed in the text.}
\label{fig:rmse}
\end{figure}

To assess the accuarcy as well as the limits of the paraxial approximation
upon increasing $|z_{\rm D}|$, we simulated OA signals at various points in the
range \mbox{$z_{\rm D}=-0.05,\ldots,-2.0 ~{\rm cm}$}. 
As pointed out in section \ref{subsubsect:Numerics_OAPoissonInt_fwd}, the Poisson 
integral based forward
solver yields an OA signal up to an amplitude factor. Hence, in order to be
able to quantitatively compare the Volterra based reconstruction $\tilde{p}_{\rm 0}$ to
the true underlying $p_{\rm 0}$, we need to adjust the respective signal
amplitudes. Therefore, in a preprocessing step we first
normalized the initial stress profile so that $\int\!p_{\rm 0}(\tau)~\mathrm{d}\tau
\equiv 1$ and subsequently adjusted the amplitude of $\tilde{p}_{\rm 0}(\tau)$ so that the
residual sum of squares ${\rm RRS}(\Delta \tau) = \sum_i^{M(\Delta \tau)}
[p_{\rm 0}(\tau_i)-\tilde{p}_{\rm 0}(\tau_i)]^2$ is
minimized within the tuning-interval $\Delta \tau = [\tau_{-}:\tau_{+}]$
containing $M(\Delta \tau)$ sampling points.
The resulting root-mean-square error ${\rm RMSE}(\Delta \tau)=\sqrt{{\rm RSS}(\Delta
\tau)/M(\Delta \tau)}$ as function of the detector-to-layer distance $|z_{\rm
D}|$ is shown in Fig.\ \ref{fig:rmse}. As evident from the figure, if the
tuning interval encloses the region around the signal edges where the
reconstructed stress profile is gently inclined, see the curve
corresponding to $\Delta z \equiv c \Delta \tau = [-0.02:0.12]$ in Fig.\ \ref{fig:rmse}, the
RMSE above $-z_{\rm D}\approx 0.2~{\rm cm}$ saturates at a finite value
characterizing the signal mismatch around the edges. In contrast, if the
tuning interval is chosen to be more narrow and to exclude those
edge-mismatches, e.g.\ in the range $\Delta z = [0.02:0.08]$, the RMSE
decreases as $\propto |z_{\rm D}|^{-3/2}$ until a limiting point at $z_{\rm
D}\approx -1.5~{\rm cm}$ is reached.  This limiting point is solely due to the
mesh-width used to discretize the $z$-axis. Here, we
considered a mesh width of $dz=2.5~{\rm \mu m}$. Note that a smaller width $dz$ would
in turn allow to shift the limiting point to larger values of $|z_{\rm D}|$ and
to obtain highly precise reconstructed stress profiles also in the deep FF.

\subsection{Inversion via successive approximations}
\label{sect:NumEx_picardIt}

\begin{figure}[t!]
\begin{center}
\includegraphics[width=0.49\linewidth]{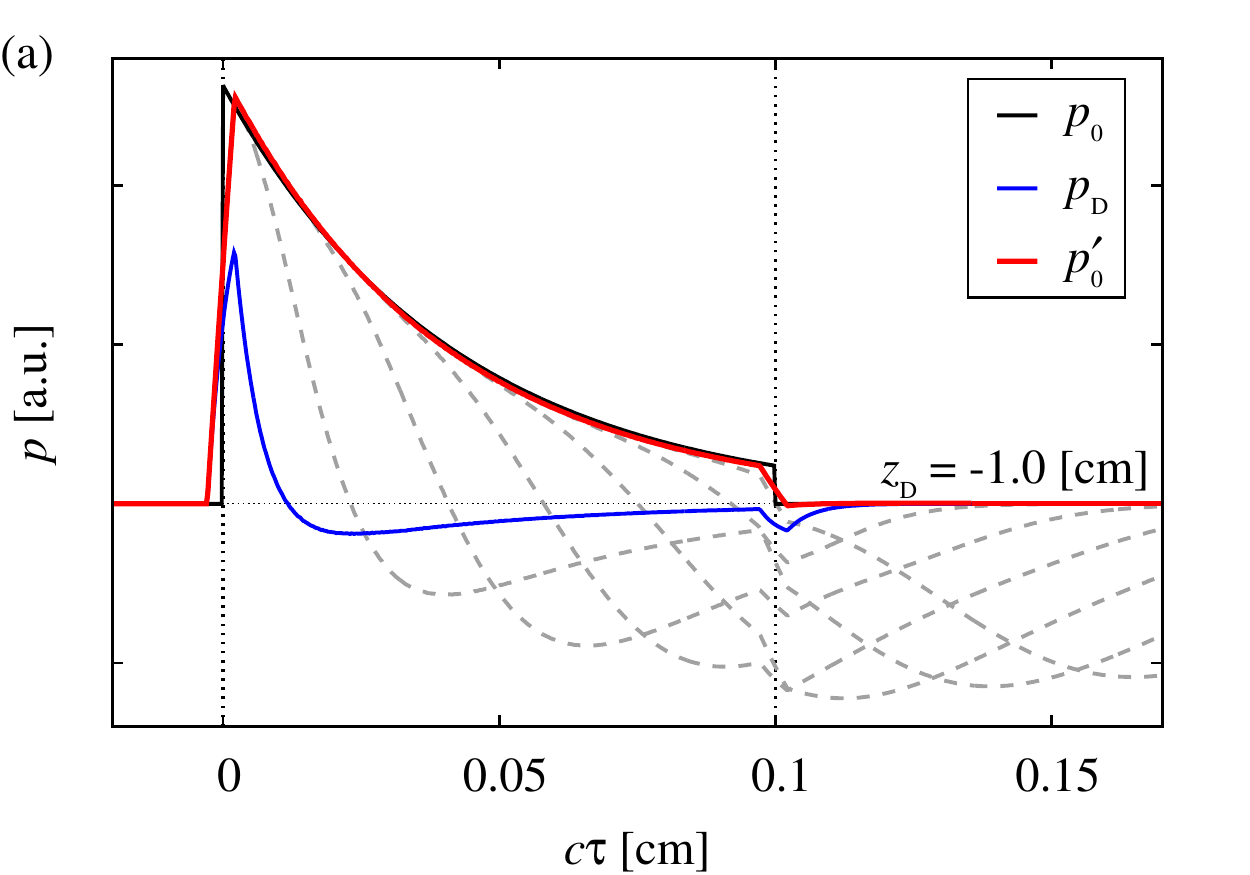}
\includegraphics[width=0.49\linewidth]{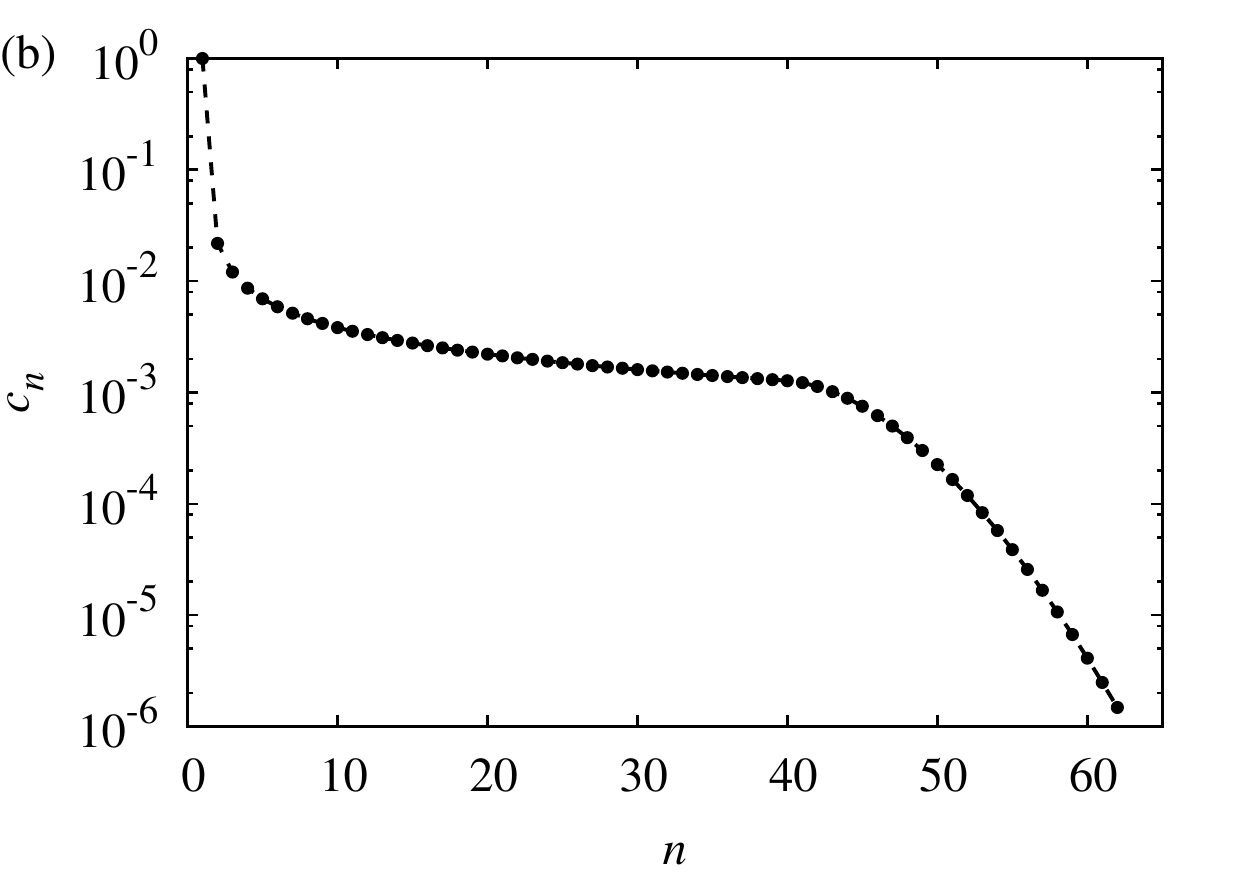}
\end{center}
\caption{
(Color online) Solution of the source reconstruction problem using a Picard-Lindel\"of
iteration scheme for synthetic input data, starting off from the low level
predictor $p_{\rm 0}^{(0)}(\tau)=0$. 
(a) Inversion of an OA signal $p_{\rm D}$, 
computed for the detection point $z_{\rm D}=-1~{\rm cm}$ (solid black solid line), 
to an initial stress profile $p_{\rm 0}^\prime$ (solid gray line). As evident
from the figure, the inverted signal compares well to the initial acoustic 
stress profile $p_{\rm 0}$. The intermediate auxiliary stress profiles
(dashed gray lines; referring to $n=5$, $10$, $15$, $20$, $25$ and $30$ iteration
cycles) feature a pronounced rarefaction dip that shifts towards larger values of
$c\tau$ upon iteration.
(b) Illustration of the convergence of the iteration scheme. The iteration
procedure is stopped as soon as the Chebychev-norm between two subsequent
profiles decreases below $c_n\equiv||p_{\rm 0}^{(n+1)}-p_{\rm 0}^{(n)}||< 10^{-6}$.
The final predictor is thus $p_{\rm 0}^\prime = p_{\rm 0}^{(n=62)}$.
\label{fig:picardIt}}
\end{figure}

An exemplary inversion of an OA signal, obtained using the Poisson integral
based forward solver in the acoustic FF at $z_{\rm D}=-1~{\rm cm}$, via the
Picard-Lindel\"of iteration scheme is illustrated in Fig.\ \ref{fig:picardIt}.
Therein, the iteration procedure was started off from a low level predictor
with $p_{\rm 0}^{(0)}(\tau)=0$.  As evident from Fig.\ \ref{fig:picardIt}(a),
the intermediate auxiliary stress profiles $p_{\rm 0}^{(n)}(\tau)$ approach
the true initial stress $p_{\rm 0}(\tau)$ upon iteration, featuring a pronounced
rarefaction dip that shifts towards larger values of $c\tau$ for increasing
$n$.  The convergence of the iteration scheme for the above OA signal is
illustrated in Fig.\ \ref{fig:picardIt}(b), where the evolution of the
Chebychev-norm $c_n\equiv||p_{\rm 0}^{(n+1)}-p_{\rm 0}^{(n)}||$ is shown. The
iteration procedure was terminated as soon as the latter decreased below the
threshold $c_n \leq 10^{-6}$. The flattening of the $c_n$ curve in the range
$n=10$ through $40$ iterations is mainly due to the pronounced rarefaction
phase featured by the auxiliary stress profiles. As soon as the latter shifts
beyond $c \tau = 0.20~{\rm cm}$, i.e.\ the boundary of the computational domain used in
our simulation, the value of $c_n$ continues to decrease noticeably,
approaching the final predictor $p_{\rm 0}^\prime = p_{\rm 0}^{(n=62)}$.

\section{Summary}
\label{sect:Summary}

In the presented article, we discussed OA signal generation in the 
paraxial approximation to the full wave equation.  
We introduced and detailed numerical schemes to simulate OA signals
in the forward direction, where we considered a Poisson integral 
based solver for the forward direction and a Volterra integral based
solver for the paraxial approximation. Further, regarding the inverse
OA problem, we considered solvers based on the OA Volterra integral, only.

By means of numerical experiments, geared towards actual laboratory
experiments for existing polyvinyl alcohol hydrogel (PVA-H) based tissue
phantoms reported in Ref.~\cite{Blumenroether:2016} we characterized OA signals in
the acoustic near- and far-field in the paraxial approximation in the forward
and inverse direction. Further, in the far-field, where the particular
theoretical model can be expected to approximate the full OA wave equation well, we
accomplished the inversion of OA signals to initial stress profiles resulting
from input obtained by solving the direct problem for the full OA wave
equation.  Thus, even if the signal is not produced by the Volterra integral
itself, the initial absorption profile can nevertheless be reconstructed using
the developed numerical procedure and, material characteristics, as, e.g., the
thickness of absorbing layers can be determined reliably. 
We further assessed the quantitative agreement between the reconstructed and
true initial stress profiles as one proceeds towards the far-field. This regime
is of particular interest since it allows for OA depth profiling for absorbing
structures. This is of pivotal relevance for various applications that strive
to reconstruct internal material properties on the basis of external OA
signal measurements.

Finally, from a point of view of computational theoretical physics, we
presented a self-contained numerical approach to the solution of the source
reconstruction problem in the field of inverse optoacoustics.  Albeit there
exist several general schemes for the forward and inverse solution of the
underlying Volterra equation, this particular inverse problem for the paraxial
approximation of the OA wave equation has not yet received much attention in
the literature.  Here, capitalizing on the particular structure of the
Volterra operator, i.e.\ the diffraction-term in the optoacoustic Volterra
integral equation, we could derive highly (time-)efficient numerical forward
and inverse solvers, thus presenting a methodological progress in the field of
optoacoustics.

It is now intriguing to conjecture a further inverse problem related to the OA
Volterra integral equation, that is, the Volterra kernel reconstruction
problem.  It aims at effectively modeling the diffraction transformation of OA
signals based on the knowledge of initial stress profiles and detected OA
signals, and is not satisfactorily solved in the literature, yet.  Such
investigations are underway in our group.

\section*{Acknowledgments}
We thank E.\ Blumenr\"other for valuable discussions and comments, as well as
for critically reading the manuscript.  J.\ S.\ acknowledges support from the
German Federal Ministry of Education and Research (BMBF) in the framework of
the project MeDiOO (Grant FKZ 03V0826).  O.\ M.\ acknowledges support from the
VolkswagenStiftung within the ``Nieders\"achsisches Vorab'' program in the
framework of the project ``Hybrid Numerical Optics - HYMNOS''  (Grant ZN 3061).
Further valuable discussions within the collaboration of projects MeDiOO and
HYMNOS at HOT are gratefully acknowledged.

\section*{References}
\bibliography{masterBibfile_optoacoustics,commentsBibfile_optoacoustics}

\begin{thebibliography}{10}
\bibitem{comment:StressConfinement}
Note that formally, the temporal delta-function might be represented by a
  sequence of functions $\delta_{t_{\rm p}}(t) = (\pi t_{\rm
  p})^{-1/2}\exp{-(t/t_{\rm p})^2}$ in the limit $\delta(t)=\lim_{t_{\rm p}\to
  0} \delta_{t_{\rm p}}(t)$. If we assume a hypothetical situation with
  absorption lengths of $\ell \approx 1~{\rm mm}$ and sound velocity
  $c=1500~{\rm m/s}$ this yields typical propagation times of $\approx 1~{\rm
  \mu s}$. Thus, pulse durations of $t_{\rm p}\approx 10~{\rm ns}$ might
  suffice to approve stress confinement.
\bibitem{comment:PrincipalValue}
Note that, so as to exclude the nonphysical effect of a point source on itself,
  the Poisson integral has to be understood in terms of its principal value if
  the source volume $V^\prime$ contains the field point $\vec{r}$.
\end{thebibliography}

\end{document}